\documentclass[11pt,oneside,letterpaper]{article}
\pdfoutput=1
\usepackage{amssymb}
\usepackage{amsmath}
\usepackage[dvips]{graphicx}
\usepackage{setspace}
\usepackage{amsfonts}
\usepackage{fancyhdr}
\usepackage{xcolor}
\usepackage{graphicx}
\usepackage{rotating}
\usepackage{comment}
\usepackage{color}
\usepackage{cite}
\usepackage{braket}

\usepackage{tikz}

\usepackage{subfigure}
\usetikzlibrary{decorations.pathmorphing}

\definecolor{darkgreen}{rgb}{0,0.5,0}
\definecolor{darkblue}{rgb}{0,0,0.6}
\definecolor{purple}{rgb}{0.4,.2,0.7}

\newcommand{\be}{\begin{equation}}
\newcommand{\ee}{\end{equation}}

\usepackage[colorlinks=true,citecolor=darkgreen,linkcolor=black,urlcolor=purple,pagebackref]{hyperref}

\usepackage{pdfsync}

\makeatletter
\newcommand*{\defeq}{\mathrel{\rlap{%
                     \raisebox{0.3ex}{$\m@th\cdot$}}%
                     \raisebox{-0.3ex}{$\m@th\cdot$}}%
                     =} 
\makeatother

\def\be{\begin{eqnarray}}
\def\ee{\end{eqnarray}}

\newcommand{\tr}{\textrm{Tr}\,}

\newcommand{\bea}{\begin{eqnarray}}
\newcommand{\eea}{\end{eqnarray}}

\newcommand{\beg}{\begin{equation} \begin{gathered}}
\newcommand{\eeg}{\end{gathered} \end{equation}}

\let\G=\Gamma
\let\l=\left
\let\r=\right

\let\d=\partial
\def\be{\begin{equation}}
\def\ee{\end{equation}}
\def\ba{\begin{array}}
\def\ea{\end{array}}

\def\eps{\epsilon}

\def \G{\Gamma}
\def \ddt{{d\over 2}}

\def\ba#1\ea{\begin{align}#1\end{align}}
\def\bs#1\es{\begin{split}#1\end{split}}

\allowdisplaybreaks  

\numberwithin{equation}{section}

\def \be {\begin{equation}}
\def \ee {\end{equation}}

\def\comma{\,,}
\def\period{\,.}

\begin{document}
\onehalfspacing

\begin{center}

~
\vskip5mm

{\LARGE  {
Cosmological Correlators at Finite Coupling
}}

\vskip10mm

Lorenzo Di Pietro$^{1,2}$, \ Victor Gorbenko$^{3}$,\ \ Shota Komatsu,$^{4}$ 

\vskip15mm

{\it $^{1}$ Dipartimento di Fisica, Universit\'a di Trieste,
Strada Costiera 11, I-34151 Trieste, Italy}\\
{\it $^{2}$ INFN, Sezione di Trieste, Via Valerio 2, I-34127 Trieste, Italy}\\
{\it $^{3}$ Laboratory for Theoretical Fundamental Physics, EPFL, Rte de la Sorge, Lausanne } \\
{\it $^{4}$ Department of Theoretical Physics, CERN, 1211 Meyrin, Switzerland}
\vskip5mm

\vskip5mm

\end{center}

\vspace{4mm}

\begin{abstract}
\noindent

We study finite-coupling effects of QFT on a rigid de Sitter (dS) background taking the $O(N)$ vector model at large $N$ as a solvable example. Extending standard large $N$ techniques to the dS background, we analyze the phase structure and late-time four-point functions. Explicit computations reveal that the spontaneous breaking of continuous symmetries is prohibited due to strong IR effects, akin to flat two-dimensional space. Resumming loop diagrams, we compute the late-time four-point functions of vector fields at large $N$, demonstrating that their spectral density is meromorphic in the spectral plane and positive along the principal series. These results offer highly nontrivial checks of unitarity and analyticity for cosmological correlators.

 \end{abstract}

\pagebreak
\pagestyle{plain}
\setcounter{tocdepth}{2}
{}
\vfill
{ \tableofcontents}

\newpage


\section{Introduction}
As far as we know it, the universe has been expanding. Unfortunately, our theoretical tools for studying non-static expanding spacetimes are not as well-developed as we would like to. What is lacking is not only a non-perturbative definition of gravity, but also our understanding of quantum fields on such backgrounds. A convenient model of an expanding space-time is de Sitter space (dS). It benefits form having the maximal possible symmetry and also it makes the expansion the most manifest, since it expands in the fastest possible way, that is exponentially. This, in turn, leads to the presence of many peculiar features of its geometry, such as a cosmological horizon.

Some progress on understanding QFT on dS was made in our recent paper \cite{DiPietro:2021sjt}, where we focused on perturbative weakly coupled theories (for other related developments see  \cite{Hogervorst:2021uvp,Sleight:2020obc, Sleight:2021plv,Loparco:2023rug,Chowdhury:2023arc,Meltzer:2021zin,Salcedo:2022aal,Jazayeri:2021fvk,Baumann:2021fxj}). In this paper we make the next step and analyze strongly-coupled theories. As a test ground we pick a very popular example of a strongly coupled theory which can in many cases be studied analytically: the $O(N)$ vector model at large $N$. Of course, it is not the most generic strongly-coupled QFT, however, it does allow us to study several phenomena in a qualitatively new regime. This includes: spontaneous symmetry breaking, unitarity and positivity of the spectral density, analytic properties of correlators as well the issue of IR divergences.

Let us briefly summarize the results related to the phenomena we just mentioned:

{\it{Symmetry breaking:}} In all situations that we observed there is no spontaneous symmetry breaking in dS. This is true even at infinite $N$ where one could expect the phenomenon to occur even in a finite volume system. In a regime where the scale of would-be symmetry breaking in flat space is much higher than Hubble, as suggested long time ago 
\cite{Ford:1985qh}, the mechanism of symmetry restoration is similar to that in two dimensions: would-be Goldstone bosons spread over the vacuum manifold restoring the symmetry and excitations behave as gapped (correlation functions decay exponentially in proper time). 

{\it{Unitarity:}} The statement of unitarity was formulated at the nonperturbative level in \cite{DiPietro:2021sjt, Hogervorst:2021uvp}: the decomposition of correlation functions with respect to the irreducible representations of the dS symmetry group $SO(1,d+1)$ must be non-negative. Here we checked this statement explicitly for the $O(N)$ model at any coupling. The check is more non-trivial than in the weakly coupled cases studied before since the disconnected contribution, which is manifestly positive, is not parametrically larger than the connected piece, which on its own can be negative.

{\it{Analyticity:}} The analytic properties of the spectral density with respect to principal series decomposition are still conjectural beyond perturbation theory. In the $O(N)$ model it is meromorphic with poles that can be associated with free field composite operators at the weak coupling, and that acquire order one shifts at strong coupling, but still stay in the ``allowed'' region, defined below.

{\it{IR effects:}} IR effects in dS are famously subtle when light fields are present in the theory. The large-$N$ expansion allows us to study such theories at finite coupling, which is a qualitatively new regime with respect to the standard (stochastic) approach \cite{Starobinsky:1986fx,Gorbenko:2019rza}. We find that the theory is stable, and for small coupling find the agreement with the results of \cite{Gorbenko:2019rza} at the subleading order in the coupling.

We are not the first to study the large-$N$ $O(N)$ model in dS, previous studies mostly motivated by the interest in the strong IR effects mentioned above. As far as we know, however, all the works considered small coupling regime \cite{Serreau:2011fu,Serreau:2013psa}, while going sometimes to the subleading order in $1/N$ \cite{Gautier:2015pca,LopezNacir:2018xto,LopezNacir:2019ord,Moreau:2020gib}. We are also aware only about the previous results on one- and two-point functions, while the late-time four-point function computation appears here for the first time. 

The rest of the paper is organized as follows. In {\bf section} \ref{sec:EffectivePotential}, we study the phase structure of the $O(N)$ model in dS by solving the gap equation and demonstrate the absence of symmetry breaking. In {\bf section} \ref{sec:2pt}, we compute the bulk two-point function of the Hubbard-Stratonovich field $\sigma$ by resumming the bubble diagrams and study its analytic properties. We show that the decay exponents of the two-point functions at weak coupling reproduce the result from the stochastic approach. In {\bf section} \ref{sec:4pt}, we compute the late-time four-point function of the $O(N)$ vector fields $\phi^{i}$ and verify the positivity and the meromorphicity of the spectral representation. Finally in {\bf section} \ref{sec:conclusion}, we discuss several promising future directions. In {\bf Appendix} \ref{ap:bubble}, we explain the computation of the bubble diagram for $dS_3$.

\section{Effective Potential and Absence of Symmetry Breaking}\label{sec:EffectivePotential}
Wick rotated dS$_{d+1}$ is a sphere, and unlike for other maximally symmetric spacetimes, Euclidean dS space is compact, which in some sense makes its study more straightforward. A harder part is often to infer the real time physics from Euclidean calculations. In our case, we will be able to extract most of the observables we need in this way. We will also make some comparison to the direct Lorentzian calculations in the case of small masses of the physical degrees of freedom. 

The Euclidean Lagrangian of the $O(N)$ model is
\begin{equation}
\mathcal{L} = \frac12 (\partial \phi^i)^2 + \frac{m^2}{2} (\phi^i)^2 + \frac{\lambda}{2 N} ((\phi^i)^2)^2~,
\end{equation}
where $i = 1, \dots, N$ and summation over $i$ is implicit. It is rather straightforward to generalize our computations to more general $O(N)$ invariant potentials, but it would not add much conceptually new so we will stick to the simplest interacting potential.

A standard way to compute the $1/N$ perturbation theory with fixed $\lambda$ is by introducing a Hubbard-Stratonovich auxiliary field $\sigma$
\begin{equation}\label{eq:HSLag}
\mathcal{L} = \frac12 (\partial \phi^i)^2 + \frac{m^2}{2} (\phi^i)^2 - \frac{1}{2 \lambda} \sigma^2 + \frac{1}{\sqrt{N}}  \,\sigma (\phi^i)^2 ~.
\end{equation}
The integration contour for $\sigma$ runs on the imaginary axis. Note that the equation of motion simply sets
\begin{equation}
\sigma = \frac{\lambda}{\sqrt{N}}(\phi^i)^2~,
\end{equation}
hence $\sigma$ is identified with the composite operator $(\phi^i)^2$ inside correlation functions.

Expanding the fields around constant values
\begin{align}
\sigma & = \sqrt{N} \Sigma + \hat{\sigma}~, \label{eq:expS}\\
\phi^i & = \sqrt{N} \Phi^i + \hat{\phi}^i~, \label{eq:expC}
\end{align}
the effective potential is the lowest order in the corresponding expansion of the Lagrangian
\begin{equation}
V(M^2, \Phi^i) = N\left(- \frac{(M^2 - m^2)^2}{8\lambda} + \frac{M^2}{2} (\Phi^i)^2 + \frac12 \tr\log\left(-\square + M^2\right)\right)~,
\end{equation}
where we are using the shifted variable $M^2 = m^2 + 2\Sigma$. Note that $M^2$ is the physical mass of the fluctuations $\hat \phi^i$ at leading order.

The equation for the vacuum at leading order at large $N$ are
\begin{align}
0 & =  \frac{\partial V}{\partial \Phi^i} =N M^2 \Phi^i  ~, \label{eq:vacC} \\
0 & = \frac{\partial V}{\partial M^2}=\frac{N}{2}\left( \frac{m^2 - M^2}{2 \lambda}  + (\Phi^i)^2 + \tr \frac{1}{-\square + M^2} \right)~.\label{eq:vacS}
\end{align}

We will now specify these formulas to the sphere $S^{d+1}$.\footnote{For studies of different observables in  the $O(N)$ model on the sphere using related techniques see \cite{Klebanov:2011gs,Giombi:2019upv}.} Equation \eqref{eq:vacC} is easily solved, the two-branches are either $\Phi^i = 0$ and any $M^2$ (no symmetry breaking) or $M^2 = 0$ and $\Phi^i\neq 0$ (symmetry breaking). At this point it appears that both solutions might be allowed, as is the case in the $O(N)$ model on AdS \cite{Carmi:2018qzm}. To write down more explicitly the second equation, we need to evaluate the trace on the sphere. We can compute the trace in a basis of eigenvalues of the Laplacian. The possible eigenvalues on $S^{d+1}$ are
\begin{equation}
-\frac{l(l+d)}{L^2}~,
\end{equation}
where $L$ is the radius and $l$ is a non-negative integer. The corresponding eigenvectors are the harmonic functions
\begin{equation}
\Omega^{S^{d+1}}(l,\zeta) =\frac{(d+2l)\Gamma (d+l)}{\Gamma (l+1) (4 \pi )^{\frac{d+1}{2}}\Gamma \left(\frac{d+1}{2}\right)}    \, _2{F}_1\left(-l,d+l;\tfrac{d+1}{2};\tfrac{\zeta }{4}\right)~,\label{eq:HarmNorm}
\end{equation}
where $\zeta\in[0,4]$ is the square of the chordal distance in units of $L^2$, and whose normalization has been chosen in such a way that they behave simply under convolution
\begin{equation}
\int_{S^{d+1}} \sqrt{g(x)}\,\Omega^{S^{d+1}}(l,x,y) \Omega^{S^{d+1}}(m,x,z) = L^{d+1}  \delta_{lm} \,\Omega^{S^{d+1}}(l,y,z)~,\label{eq:Omegaconv}
\end{equation}
and that they satisfy the orthogonality relation
\begin{equation}
\int_{S^{d+1}} \sqrt{g(x)}\,\Omega^{S^{d+1}}(l,x,y) \Omega^{S^{d+1}}(m,x,y) = L^{d+1} \frac{(d+2 l)  \Gamma (d+l)}{ \Gamma (l+1)(4 \pi )^{\frac{d+1}{2}}\Gamma \left(\frac{d+1}{2}\right)} \delta_{lm}~.\label{eq:Omegaorth}
\end{equation}
With this normalization, the propagator that satisfies
\begin{equation}
(-\square_{S^{d+1}}+ M^2 )G^{S^{d+1}}(x,y) = \delta^{d+1}(x,y)~,\label{eq:GreenEq}
\end{equation}
is given by
\begin{equation}
G^{S^{d+1}}_\nu(\zeta) = \frac{1}{L^{d-1}}\sum_{l=0}^\infty  \frac{1}{l(l+d) +\frac{d^2}{4}+\nu^2} \Omega^{S^{d+1}}(l,\zeta)~.
\end{equation}

We chose to express the mass in terms of the parameter $\nu$ as $M^2 = L^{-2}(\frac{d^2}{4}+\nu^2)$. Picking the positive square-root this gives
\begin{equation}
\nu =  \sqrt{L^2\,M^2-\frac{d^2}{4}}~.
\end{equation}
In the range of masses $0\leq M^2 \leq L^{-2} \, \frac{d^2}{4}$ the parameter $\nu$ lies in the interval $i[-\frac{d}{2},0]$ of the imaginary axis, while for $M^2 > L^{-2}\, \frac{d^2}{4}$ the parameter $\nu$ is real and positive. 

We get 
\begin{align}
\begin{split}
\tr \frac{1}{-\square + M^2} & = G_\nu^{S^{d+1}}(\zeta=0)\\
&  = \frac{1}{L^{d-1}}\sum_{l=0}^\infty \frac{1}{l(l+d)+\frac{d^2}{4}+\nu^2}\frac{(d+2 l)  \Gamma (d+l)}{ \Gamma (l+1)(4 \pi )^{\frac{d+1}{2}}\Gamma \left(\frac{d+1}{2}\right)}\\
& =\frac{1}{L^{d-1}} \frac{\sec \left(\frac{\pi  d}{2}\right) \cosh (\pi  \nu ) \Gamma \left(\frac{d}{2}+i \nu \right) \Gamma \left(\frac{d}{2}-i \nu \right)}{(4 \pi )^{\frac{d+1}{2}} \Gamma \left(\frac{d+1}{2}\right)}~.
\end{split}
\end{align}
In the last line we performed the summation assuming $d<1$ and then analytically continued in $d$, i.e. we regulated the UV divergence using dimensional regularization. This divergence just comes from the bulk tadpole diagram, which indeed has a UV divergence when there are at least 2 bulk dimensions i.e. $d\geq 1$. Note that the combination
\begin{equation}
\frac{m^2}{2\lambda} + \tr \frac{1}{-\square + M^2}~,
\end{equation}
appearing in \eqref{eq:vacS} is actually finite, so the end result for the physical mass $M^2$ is scheme-independent as it should. The dimreg result around the integer dimensions $d=1,2,3$ is as follows (with $L=1$)
\begin{align}
\begin{split}
&d=1:~~ -\frac{1}{2 \pi  (d-1)}-\frac{\psi\left(i \nu +\frac{1}{2}\right)+\psi\left(\frac{1}{2}-i \nu \right)+\gamma -\log (4 \pi )}{4 \pi }~,\\
&d=2:~~ -\frac{\nu  \coth (\pi  \nu )}{4 \pi }~,\\
&d=3:~~ \frac{4 \nu ^2+1}{32 \pi ^2 (d-3)}+\frac{\left(4 \nu ^2+1\right) \left(\psi\left(i \nu +\frac{3}{2}\right)+\psi\left(\frac{3}{2}-i \nu \right)+\gamma -1-\log (4 \pi )\right)}{64 \pi ^2}~,\\
\end{split}
\end{align}
where $\psi$ is the digamma function and $\gamma$ is the Euler-Mascheroni constant. Note that dimensionally-regularized trace has a pole in any odd $d$ but is finite in any even $d$ (matching the expectation for a $d+1$ dimensional bulk).

The resulting function diverges in the limit $M^2 \to 0$, or equivalently $\nu\to -i \tfrac{d}{2}$, as follows
\begin{equation}
\tr \frac{1}{-\square + M^2} \underset{M^2\to 0}{=} ~  \frac{\Gamma \left(\frac{d}{2}\right)}{4 \pi ^{\frac{d}{2}+1} \left(\frac{d}{2}-i \nu \right)} +\mathcal{O}\left((\tfrac{d}{2}-i\nu)^0\right)~.
\end{equation}
This is simply the contribution from the $l=0$ mode in the sum. Due to this divergence, $M^2 = 0$ cannot be a solution of \eqref{eq:vacS}, because the rest of the equation stays finite in the limit.\footnote{Actually, we could also consider the possibility that $\Phi^i$ diverges like $\frac{1}{\sqrt{M^2}}$ in the limit. This solves both equations in the limit $M^2\to 0$. However the signs in eq. \eqref{eq:vacS} are such that one would need a purely imaginary coefficient of $\frac{1}{\sqrt{M^2}}$ in $\Phi^i$. It seems unphysical to have a divergent vacuum expectation value, and the imaginary coefficient would violate unitarity. Therefore we will not consider this possibility further.} This is the simple way in which the impossibility of symmetry breaking manifests itself at large $N$. 

\begin{figure}
\includegraphics[width=10cm]{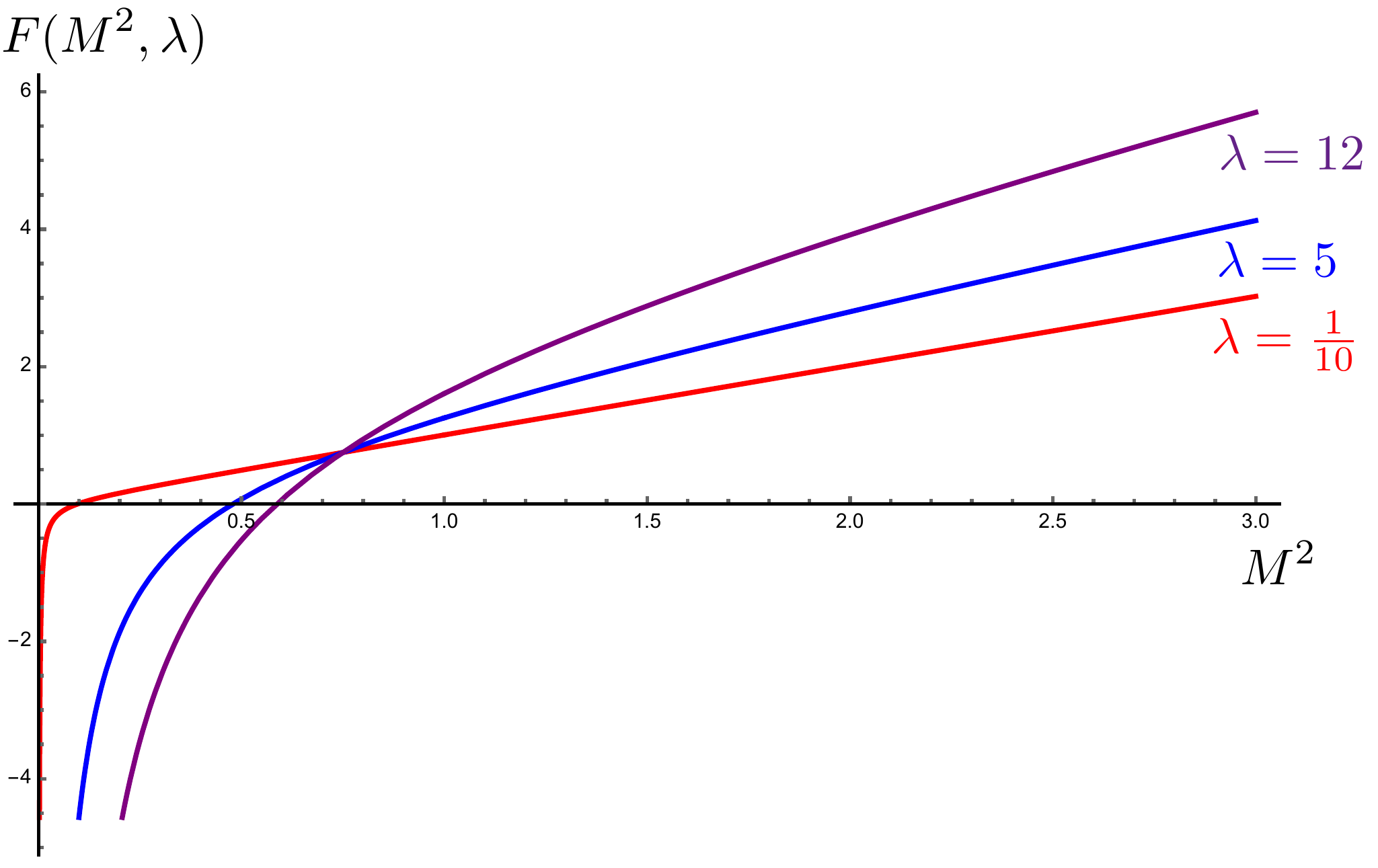}
\centering
\caption{Plot of $F(M^2, \lambda)$ as a function of $M^2\geq 0$ for $L=1$ and three different values of $\lambda$.}\label{fig:Fplot}
\end{figure}

Let us study what is the vacuum solution for the physical $M^2$ as a function of the parameters $m^2$ and $\lambda$ in the Lagrangian. Note that $\lambda$ does not need a renormalization for $d<3$ so in this range we can take the positive parameter $\lambda > 0$ in the Lagrangian to be the physical value, while $m^2$ needs a renormalization for any $d\geq 1$ as discussed above. For definiteness we consider $d=2$, i.e. dS$_3$. Working in dimreg the real parameter $m^2$ in the Lagrangian does not need any subtraction. We need to solve
\begin{align}
\begin{split}
F(M^2, \lambda) & = m^2~,\\
\text{where}~F(M^2, \lambda) & \equiv M^2 + \lambda \frac{\sqrt{1-L^2 M^2} \cot \left(\pi  \sqrt{1-L^2 M^2}\right)}{2 \pi L }~.
\end{split}
\label{Msol}
\end{align} 
Obviously turning off the interaction $\lambda = 0$ we have the solution $m^2 = M^2$. We note that for any $\lambda$ the function $F(M^2,\lambda)$ is a monotonically increasing function of $M^2\geq 0$, whose graph passes through the point $(\frac{3}{4 L^2},\frac{3}{4 L^2})$, and goes to $-\infty$ in the limit $M^2 \to 0^+$. Unlike what might appear from the expression, nothing special happens for $M^2 = \frac{1}{L^2}$, in the range $M^2 > \frac{1}{L^2}$ perhaps the function is best thought by replacing $\sqrt{1-L^2 M^2} \cot \left(\pi  \sqrt{1-L^2 M^2}\right) \to \sqrt{L^2 M^2 - 1} \coth \left(\pi  \sqrt{L^2 M^2-1}\right) $. In the limit of large $M^2$ we recover the flat-space result
\begin{equation}
F(M^2, \lambda)\underset{M^2 \to \infty}{=}~ M^2 +\lambda \frac{\sqrt{M^2}}{2\pi}(1+\mathcal{O}(M^{-2}))~.
\end{equation}
As a result, for any value of $m^2\in\mathbb{R}$ we find a solution $M^2 >0$ for the physical mass-squared. In fig. \ref{fig:Fplot} we show the plot of $F(M^2, \lambda)$.

Note that as a consequence of the interaction, for any $\lambda > 0$ the excitations are gapped, irrespectively of $m^2$. This is again the statement of the absence of symmetry breaking. Let us discuss this issue in a few more details.
\paragraph{Absence of symmetry breaking in dS.}
Generally we expect that spontaneous symmetry in dS never occurs \cite{Ford:1985qh,Ratra:1984yq}, at least for compact internal symmetry groups.
There are two lines of reasoning that lead to this statement. First, is that the Euclidean version of dS is a sphere, a finite volume manifold, on which there is no symmetry breaking. Even though some infinite-$N$ theories exhibit a phase transition even in the finite volume \cite{Russo:2019lgq}, we do not expect it to occur in the $O(N)$ model in question. One may be worried that since in real time dS expands exponentially fast, or even has infinite volume if considered in Poincar\'e coordinates, the finite volume argument is somehow invalid. To remove this suspicion, there is also a real time intuition that we present below. 

The second line of reasoning, first emphasized in \cite{Ford:1985qh}, that leads also to the absence of symmetry breaking is the following: the would-be Goldstone bosons of a broken symmetry would have logarithmic two-point functions leading to IR divergences and strong coupling at low energies. They thus spread out along the vacuum manifold dynamically restoring the symmetry, similarly to what happens in two-dimensional flat space \cite{Coleman:1973ci, Mermin:1966fe}.\footnote{This argument is related to the discussion of IR divergences for light scalar fields in dS we briefly mention below. In particular, the only mechanism to produce a protected scalar field apparently requires a spontaneous breaking of a space-time symmetry, for example having a dS brane in some ambient space with more isometries. Then the embedding coordinates of the brane will be the goldstone fields of the isometries broken by the brane and such goldstones will be protected form mass generation. See \cite{Bonifacio:2018zex} for some examples of scalars with shift symmetries on dS.} This logic, however, does not explain why discrete symmetries do not get broken in dS. In known cases, see for example \cite{Arai:2011dd}, there is a scalar field which can be thought of as the order parameter for the symmetry, and the ground state is symmetry-preserving, even if this field is heavy. It appears that dS combines both features of finite volume and two dimensions, as far as spontaneous symmetry breaking is concerned. 

While the general arguments presented above are very reasonable, there has been several claims of spontaneous symmetry breaking in dS, see for example \cite{Boyanovsky:2012nd} and references therein. These claims were debunked in later papers, in particular the $O(N)$ example at small coupling $\lambda$ was studied using two different methods in \cite{Lazzari:2013boa} and \cite{Serreau:2013eoa} which found no symmetry breaking. Our analysis gives a confirmation that symmetry does not occur also at strong coupling. We should say that we did not consider a possibility that the ground state breaks spontaneously also dS isometries and generically we do not expect this to happen.

\section{Exact $\sigma$ Propagator}\label{sec:2pt}

Let us compute the two-point function of the field $\sigma$ at leading order at large $N$ and exactly in the coupling constant $\lambda$. This will tell us about the spectrum of $O(N)$ singlet excitations at leading order at large $N$. We will compute this spectrum as a function of two parameters that we take as inputs defining the theory, namely the quartic coupling $\lambda$ (assuming $d<3$ no renormalization is needed so this coincides with the parameter in the Lagrangian) and the physical mass $M^2$ of the $O(N)$ vector excitations.\footnote{$M^2$ is the solution to the minimization of the effective potential showed in the previous section, it is formally given by a UV divergent expression in terms of $\lambda$ and the bare mass $m^2$, which becomes a finite expression when we go to the renormalized $m^2$. However the precise definition of the renormalized $m^2$ depends on the scheme, so it has some intrinsic arbitrariness. The only thing that matters is that all values of the physical $M^2$ are obtained upon varying $m^2$. As a result we can just take $M^2$ as a free parameter.} 

The (bulk-to-bulk) propagator of $\sigma$ is obtained from the resummation of bubble diagrams. It can be expressed as
\begin{equation}\label{eq:conv}
\langle \sigma(X) \sigma(Y) \rangle =  -\lambda \sum_{n=0}^\infty(-\lambda)^n \left(\star 2B \right)^n(X,Y)~.
\end{equation}
where $\star$ denotes the convolution, and $B(X,Y)$ is the bubble function defined by the following correlator in free theory
\begin{equation}
B(X,Y) = \frac{1}{2N}\langle (\hat\phi^i\hat\phi^i)(X)(\hat\phi^j\hat\phi^j)(Y)\rangle = (G_{\nu_\phi}(X,Y))^2~.
\end{equation}
Here $G_{\nu_\phi}(X,Y)$ denotes the $dS_{d+1}$ propagator for a free scalar field of mass-squared $M^2$, labeled by the parameter $\nu_\phi = -i\sqrt{d^2/4 - L^2 M^2}$. There are various such propagators depending on the ordering of the operators. The convolution integrates the time variable over the full in-in contour, so that the ``intermediate'' bubbles are considered with all possible orderings. 

The spectral representation of the bubble 
\begin{equation}
B(X,Y) = \int_{-\infty}^{+\infty} d\nu\, \frac{\nu}{\pi i} \hat{B}(\nu)\, G_{\nu}(X,Y)~,
\end{equation}
can be computed either using the relation to EAdS$_{d+1}$ valid in perturbation theory \cite{Sleight:2020obc,DiPietro:2021sjt,Sleight:2021plv}, or the Wick rotation to the sphere. Given $\hat{B}(\nu)$, we will now explain how to obtain the spectral representation for the two-point function of $\sigma$, which is defined similarly as follows
\begin{equation}
\langle \sigma(X) \sigma(Y) \rangle = \int_{-\infty}^{+\infty} d\nu\, \frac{\nu}{\pi i} f_\sigma(\nu)\, G_{\nu}(X,Y)~.
\label{sigma2pt}
\end{equation}

Using the notation in the appendix C of \cite{DiPietro:2021sjt}, the spectral representations $f(\nu)$ of a generic two-point function on $dS_{d+1}$ and that of its Wick rotation $f^S(l)$ on S$_{d+1}$ are related by 
\begin{equation}
f(\nu) = - f^S(-\tfrac{d}{2}+i\nu)~.
\end{equation}
The spectral representation on $S^{d+1}$ maps convolutions to product. Therefore equation \eqref{eq:conv} allows us to express the spectral representation for the two-point function of sigma in terms of that of the bubble via a simple geometric resummation as follows
\begin{equation}
f_\sigma^S(l) = -\frac{1}{\frac{1}{\lambda} + 2 \hat{B}^S(l)}~ .
\end{equation}
From this we readily obtain
\begin{equation}\label{eq:fsigma2pt}
f_\sigma(\nu) = \frac{1}{\frac{1}{\lambda} - 2 \hat{B}(\nu)}~ .
\end{equation}
The poles of $f_\sigma(\nu)$ determine the powers of the conformal time that appear in the late-time limit of the two-point function, which can be thought of as dS quasi-normal modes, see the discussion in \cite{DiPietro:2021sjt}. From the point of view of the CFT at the late-time boundary, these are formally scaling dimensions of $O(N)$ singlet operators. However, since in general they do not belong to unitary representations of the dS isometries, it remains unclear if one should think of them as valid operator insertions at late times.

\begin{figure}
\includegraphics[width=10cm]{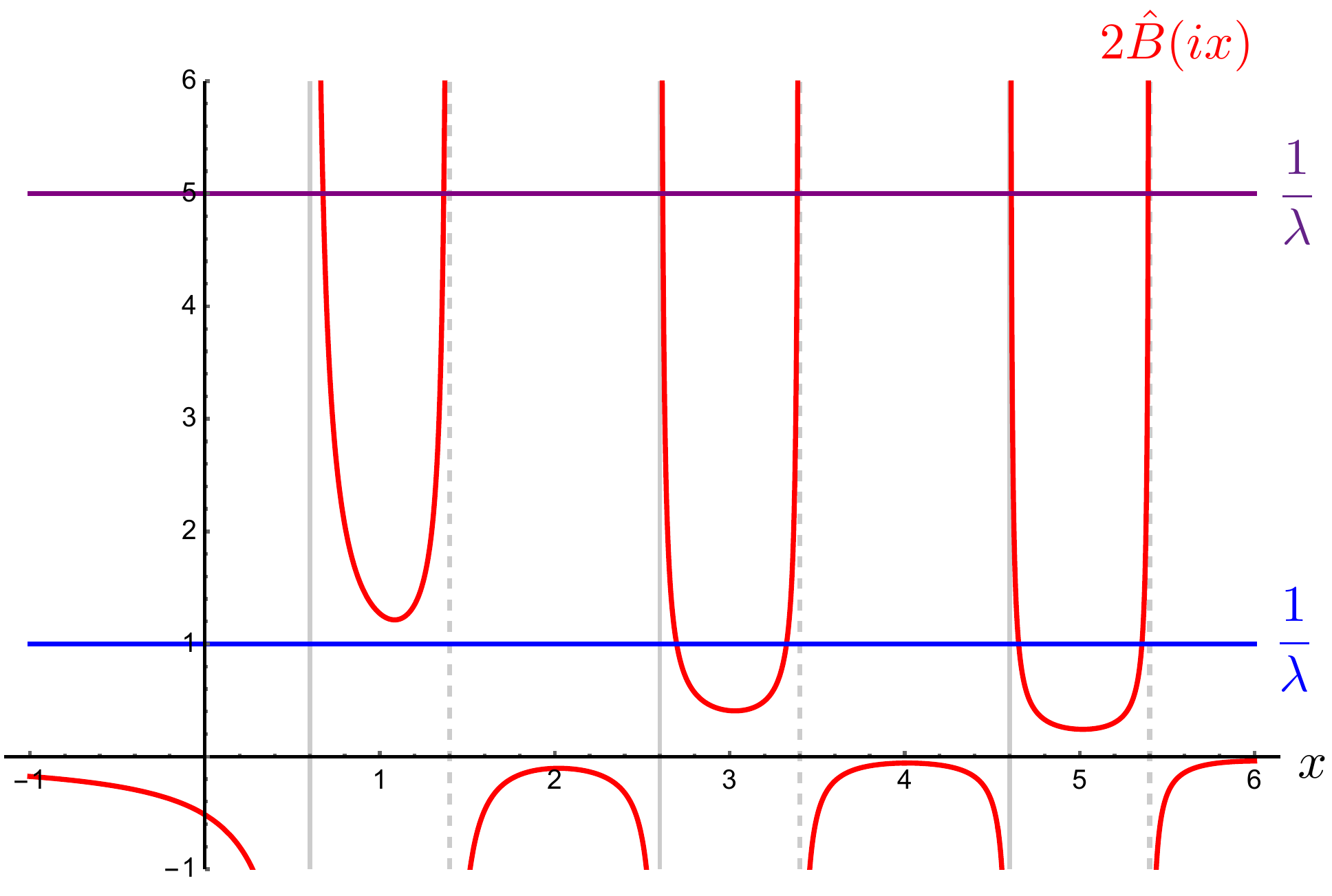}
\centering
\caption{The values of $x=-i\nu$ for which $2 \hat{B}(\nu)$ (in red) intersects the constant $\frac{1}{\lambda}$ (in purple and blue for $\lambda=\frac{1}{5}$ and $\lambda=1$, respectively) determine the dimensions $\frac{d}{2}-i\nu$ of $O(N)$-singlet operators. In figure we have taken $d=2$ i.e. dS$_3$, and $\nu_\phi = -\frac{i}{5}$. The solid vertical lines are at the values of $x$ associated to the $++$ double-trace operators, and the dashed ones are the same for the $--$ operators. We see that in this case for $\lambda =1$ the first couple of double-trace operators, i.e. those with $n=0$, have acquired an imaginary scaling dimension.}\label{fig:polesfsigma}
\end{figure}

Let us now restrict to $d=2$, i.e. dS$_3$. In this case the computation  (see Appendix \ref{ap:bubble} for details of the computation on the sphere) gives
\begin{equation}\label{eq:bubbled2}
\hat{B}(\nu) \vert_{d=2}= \frac{i}{8 \pi  \nu} \left[\pi -i \coth (\pi  \nu_\phi ) \left(\psi\left(-i \nu_\phi +\frac{i \nu}{2}+\frac{1}{2}\right)-\psi\left(i \nu_\phi +\frac{i \nu}{2}+\frac{1}{2}\right)\right)\right]~,
\end{equation}
where $\psi$ is the digamma function. We illustrate the location of the poles of $f_\sigma(\nu)$ in fig. \ref{fig:polesfsigma} by plotting the graph of the function $2 \hat{B}(\nu)$ and looking at its intersections with the constant $\frac{1}{\lambda}$. We see that at weak coupling these intersections are close to the perturbative values corresponding to the two family of double-trace operators, one with scaling dimensions $\frac{d}{2}-i \nu_{n\,++} = 2(\frac{d}{2} - i \nu_\phi) + 2n$, and another one with dimensions $\frac{d}{2}-i \nu_{n\,--} = 2(\frac{d}{2} + i \nu_\phi) + 2n$, where in both cases $n$ is a positive integer. We consider the case of a light $\hat\phi$, with $\nu_\phi\in i[-\tfrac{d}{2},0]$, for which both the scaling dimension of $\hat\phi$ and those of the double trace operators are real number. As we go towards strong coupling these operators get larger and larger anomalous dimensions, of a definite sign, namely positive for the ++ and negative for the $--$ operator. We observe that there are strong coupling thresholds $\lambda_1 < \lambda_2 < \dots$ such that at the value $\lambda_n$ the two zeroes of the denominator corresponding to $\nu_{n\,++}$ and $\nu_{n\,--}$ annihilate, causing the associate double-trace operators to get imaginary scaling dimensions.
We can expand $2\hat{B}(\nu)$ in Laurent series around the singularities, by setting $\nu = \nu_{n\pm\pm} +\delta\nu_{n\pm\pm}$ and keeping only the $1/\delta\nu_{n\pm\pm}$ term
\begin{equation}
2\hat{B}(\nu_{n\pm\pm} +\delta\nu_{n\pm\pm}) \underset{\nu\sim\nu_{n\pm\pm}}{\sim}  -\frac{ \coth (\pi  \nu_\phi )}{2 \pi  \delta\nu_{n\pm\pm}  (2 i \nu_\phi \mp (2 n+1))} + \mathcal{O}\left((\delta\nu_{n\pm\pm})^0\right)~.
\end{equation}
This expression can be then used to find a closed form expression for the shift of the double-trace pole, with $\delta\nu_{n\pm\pm}\propto\lambda$, which is valid for small $\lambda$, i.e. in perturbation theory. We get
\begin{equation}
- i \delta\nu_{n\pm\pm} = i \lambda \frac{\coth (\pi  \nu_\phi )}{2 \pi (2 i \nu_\phi \mp (2 n+1))} +\mathcal{O}(\lambda^2)~.
\end{equation}

The limit $m^2 \ll \sqrt{\lambda}L^{-\frac{3}{2}}$ and $\lambda L \ll1$ is particularly interesting because it is the one in which IR divergences appear in the standard perturbation theory. There are known ways to deal with these divergences \cite{Starobinsky:1986fx,Starobinsky:1994bd,Gorbenko:2019rza} that lead to the differential equations of the Fokker-Planck type which determine the correlation functions. This approach is historically refereed to as ``stochastic''. Let us compare our result with those obtained using this approach.

\paragraph{Comparison with the stochastic approach.}The limit $m^2 \ll \sqrt{\lambda}L^{-\frac{3}{2}}$ and $\lambda L\ll1$ for the $O(N)$ model was studied  in \cite{Gorbenko:2019rza} where, in particular, the decay exponent for  singlet and vector excitations was determined to the subleading order in $\lambda$.

Let us start with the vector excitations, $\hat{\phi}^i$ in our language. Solving \eqref{Msol} in the limit of interest we find
\begin{equation}\label{eq:IRdivvec}
M^2 = \frac{\sqrt{\lambda}}{\pi L^\frac{3}{2}}-\frac{3}{8} \frac{\lambda}{L \pi^2} +\frac{m^2}{2} + \mathcal{O}(m^2 \sqrt{\lambda}, \lambda^{\frac{3}{2}})~.
\end{equation}
The non-analiticity in $\lambda$ and the appearance of $\sqrt{\lambda}$ as a natural expansion parameter is a marker for the divergences in standard perturbation theory. The leading order agrees with the value of the exponent $\lambda_v$ found in this limit in \cite{Gorbenko:2019rza}, while in order to find agreement at the subleading order one needs to appropriately match the scheme-dependent parameter $m^2$. We leave this exercise for future work.

Let us switch to the singlet excitations.  We can use the result \eqref{eq:IRdivvec} to find that the there is a pole of the $\sigma$ two-point function at 
\begin{equation}
i\nu = 1 - 2\frac{\sqrt{\lambda}}{\pi} -\frac{3 \lambda}{2\pi^2} +\mathcal{O}(\lambda^\frac{3}{2})~, 
\end{equation}
which corresponds to a parametrically light $O(N)$-singlet mode in this limit. We see again that $\sqrt{\lambda}$ is the expansion parameter. In this case even at the subleading order there is no dependence on the scheme-dependent parameter $m^2$. The exponent  $\lambda_s$, found in \cite{Gorbenko:2019rza}  should be compared with $1-i \nu $. Indeed we find a match once we account for the difference in the definition of $\lambda$.

This result, together with \eqref{eq:IRdivvec}, is interesting for two reasons. First, we checked the computations of \cite{Gorbenko:2019rza} by a direct resummation of the diagrams, at least in the large-$N$ limit. Note that at the technical level the computation is very different. In  \cite{Gorbenko:2019rza} the exponents appear as Eigenvalues of some large-$N$ quantum-mechanical Hamiltonian. Second, together with the results of \cite{Gorbenko:2019rza}, we now see that the theories of light self-interacting scalars are under control at least in two cases: for any $N$ with a small coupling, and for large $N$ with any coupling. This provides a strong evidence for stability of generic theories of scalar fields on de Sitter. Reference \cite{Cohen:2021fzf} reported the results for exponents at the sub-subleading order in $\lambda$ for the single field. It would be nice to extend their result to the $O(N)$ model and compare with the corresponding expansion of \eqref{Msol} and \eqref{eq:IRdivvec}.

\paragraph{CFT limit.} The two-point function of $\sigma$ drastically simplifies in the limit $\lambda\to\infty$ and $\nu_{\phi}=-i/2$;
\ba
\left.f_{\sigma}(\nu)\right|_{\lambda\to\infty, \nu_{\phi}=-\frac{i}{2}}=4i\nu\period
\ea
This is the limit in which the bulk $O(N)$ model becomes conformal. Since the Poincar\'{e} patch of dS$_{d+1}$ can be mapped to a half of flat Minkowski space ($t<0$) by a Weyl transformation, we expect that the two-point function exhibits a power law in this limit. This can be verified explictly by performing the spectral integral \eqref{sigma2pt}. To do so, we use the representation of the dS two-point function in terms of ``quasi-normal modes'', i.e.~poles on the upper half plane (UHP), given in (C.20) of \cite{DiPietro:2021sjt}:
\ba\label{eq:sumresidues}
\langle \sigma (X) \sigma(Y)\rangle=\sum_{\nu_{\ast}:\text{poles on UHP}}{\rm Res}\left[\nu\rho^{ll}_{\sigma}(\nu)\right] G_{-\nu_{\ast}}^{\rm AdS}\left(\frac{4}{\zeta}\right)\comma
\ea
Here the spectral density $\rho^{ll}_{\sigma}$ and the AdS propagator $G_{-\nu}^{\rm AdS}$ are
\ba
\begin{aligned}
\rho^{ll}_{\sigma} (\nu)&\equiv \frac{i\nu\Gamma (\pm i\nu)}{2\pi} (e^{\pi \nu}f_{\sigma}(\nu)-e^{-\pi \nu}f_{\sigma}(-\nu))=4\nu\coth \pi \nu\period\\
G_{-\nu}^{\rm AdS}&=\frac{1}{2\pi (-\zeta)^{1-i\nu}}\,{}_2F_{1}\left(1-i\nu,\frac{1}{2}-i\nu,1-2i\nu,\frac{4}{\zeta}\right)\comma
\end{aligned}
\ea
and the dS invariant distance $\zeta$ (in the Poincar\'{e} coordinates) reads
\ba
\zeta=\frac{-\eta_{12}^2+|x_{12}|^2}{\eta_1\eta_2}\period
\ea
Evaluating the residues one by one and expanding the result at large $\zeta$, we find the folllowing cancellation pattern;
\ba
\begin{array}{lccccc}
\nu_{\ast}=i:\quad&-\frac{2}{\pi^2\zeta^2}&-\frac{8}{\pi^2\zeta^3}&-\frac{30}{\pi^2\zeta^4}&-\frac{112}{\pi^2\zeta^5}&-\frac{420}{\pi^2\zeta^6}+\cdots\\
\nu_{\ast}=2i:\quad&&+\frac{8}{\pi^2\zeta^3}&+\frac{48}{\pi^2\zeta^4}&+\frac{224}{\pi^2\zeta^5}&+\frac{960}{\pi^2\zeta^6}+\cdots\\
\nu_{\ast}=3i:\quad&&&-\frac{18}{\pi^2\zeta^4}&-\frac{144}{\pi^2\zeta^5}&-\frac{810}{\pi^2\zeta^6}+\cdots\\
\nu_{\ast}=4i:\quad&&&&+\frac{32}{\pi^2\zeta^5}&+\frac{320}{\pi^2\zeta^6}+\cdots\\
\nu_{\ast}=5i:\quad&&&&&-\frac{50}{\pi^2\zeta^6}+\cdots\period
\end{array}
\ea
As a result we are left with a single power $\propto \zeta^{-2}$ reproducing the expected dependence of the two-point function of $\sigma$ at the conformal point.

\section{Four-point function: unitarity and OPE structure}\label{sec:4pt}
We proceed to studying the four-point function. The calculation is very similar to the one done in \cite{DiPietro:2021sjt} and for this reason we will skip most of the details. Indeed, after the Hubbard-Stratonovich transformation we effectively study the same cubic theory, the only technical difference being a non-standard (very simple) propagator for $\sigma$. An important distinction is that now we can trust the expressions obtained for a finite coupling. We will focus on the singlet channel since it allows us to neglect the cross-channel connected diagrams at the leading non-trivial order. Let us first study the connected contribution produced by the $\sigma$-exchange diagram. We already computed the spectral decomposition for  the two-point function of $\sigma$ in \eqref{sigma2pt}. From it, following exactly the same steps as in \cite{DiPietro:2021sjt}, we get the the four-point function in the conformal partial waves decomposition form:
\beg
\label{4ptconn}
\l \langle \hat{\phi}^i(x_1,\eta_c)\hat{\phi}^j(x_2,\eta_c)\hat{\phi}^k(x_3,\eta_c)\hat{\phi}^l(x_4,\eta_c)\r \rangle_c = \\
=\delta^{ij} \delta^{kl}\l( \frac{\eta_c}{x_{12}}\r)^{d-2i\nu_\phi} \l(\frac{\eta_c}{x_{34}}\r)^{d-2i\nu_\phi} \int_{-\infty-i(\ddt-\eps)}^{+\infty-i(\ddt-\eps)} d \nu \, f^{\rm 4pt}(\nu) {\cal F^{\{-\nu_\phi\}}_\nu}(z,\bar z) \,,
\eeg
where 
\begin{align}
\begin{split}
\label{f4pt}
& f^{\rm 4pt}(\nu) = \frac{4}{N} \, A(\nu) \, f_\sigma(\nu)~,\\
 A(\nu) & \equiv \frac{\G^4(i\nu_\phi)}{2^{10}\pi^{2d+5}}\sin^2\tfrac{\pi}{2}\l(\tfrac{d}{2}-i\nu-2 i \nu_\phi\r)\G^2\l(\tfrac{d-4i\nu_\phi\pm2i\nu}{4}\r)\G^2\l(\tfrac{d\pm2i\nu}{4}\r) \frac{\nu}{\pi i}\,,
\end{split}
\end{align}
where $f_\sigma(\nu)$ is given in eq. \eqref{eq:fsigma2pt}, and ${\cal F^{\{-\nu_\phi\}}_\nu}(z,\bar z) $ is the spin zero conformal partial wave (see e.g.~\cite{Karateev:2018oml}) and $\Gamma (a\pm b)\equiv \Gamma(a+b)\Gamma(a-b)$.

We are now going to analyze the analytic structure of the answer and check the positivity condition imposed by unitarity. For both of these tasks we need to remember the disconnected four-point function in the $t$ and $u$ channels 
\beg
\label{4ptdisc}
\l\langle\hat{\phi}^i(x_1,\eta_c)\hat{\phi}^k(x_3,\eta_c)\r\rangle\l\langle\hat{\phi}^j(x_2,\eta_c)\hat{\phi}^l(x_4,\eta_c)\r\rangle+ \l\langle\hat{\phi}^i(x_1,\eta_c)\hat{\phi}^l(x_4,\eta_c)\r\rangle\l\langle\hat{\phi}^j(x_2,\eta_c)\hat{\phi}^k(x_3,\eta_c)\r\rangle\Bigg | _{J=0}= \\
=\frac{\delta^{i k} \delta^{j l}+\delta^{i l} \delta^{j k}}{2}\l( \frac{\eta_c}{x_{12}}\r)^{d-2i\nu_\phi} \l(\frac{\eta_c}{x_{34}}\r)^{d-2i\nu_\phi} \int_{-\infty}^{\infty} d\nu\,\rho^{\rm free}(\nu) {\cal F_{\nu}^{\{-\nu_\phi\}}}\,,
\eeg
with the spectral density given by \cite{DiPietro:2021sjt,Karateev:2018oml}
\beg
\label{rhofree}
\rho^{\rm free}(\nu)= \frac{\G\l(\ddt\r) \G^4\l(i\nu_\phi\r)}{64 \pi^{\frac{3 d}{2}+3}}
\frac{\G\l(\ddt\pm i\nu\r)\G\l(\frac{d-4i\nu_\phi\pm2i\nu}{4}\r)}{\G(\pm i\nu)\G\l(\frac{d+4i\nu_\phi\pm2i\nu}{4}\r)}\,.
\eeg

Now, let us consider the contribution to the singlet channel, meaning that we take $\delta_{ij} \delta_ {kl}\l \langle \hat{\phi}^i\hat{\phi}^j\hat{\phi}^k\hat{\phi}^l\r \rangle /N$. Then \eqref{4ptconn} and \eqref{4ptdisc} both contribute at $\mathcal{O}(1)$ (see a similar discussion for AdS in \cite{Carmi:2018qzm}). It thus leads us to the study of the total spectral density 
\bea
\begin{split}
\label{rhotot}
\rho(\nu)   = \rho^{\rm free}(\nu) + N\rho^c(\nu)\,,\\
\rho^c(\nu)= \frac{1}{2}(f^{\rm 4pt}(\nu) + f^{\rm 4pt}(-\nu))\,.
\end{split}
\eea

We would like to discuss the analytic properties of this function. First of all we note that it is meromorphic. This implies that by an appropriate deformation of the $\nu$-contour we can express the four-point function as a discrete sum of conformal blocks. This produces an OPE decomposition of our four-point correlator. 
 We also expect that this OPE decomposition does not produce any new operators, beyond the three series of operators that at weak coupling can be labeled as 
 \be
O^i\d^{2n}O^i,\qquad O^i\d^{2n}\tilde O^i,\, \text{and }\qquad \tilde O^i\d^{2n}\tilde O^i\,,
\ee
where $O^i$ and $\tilde O^i$ stand for the two modes of the free field $\hat\phi^i$ that behave as $-\eta^{\frac{d}{2}+i\nu_\phi}$ and $-\eta^{\frac{d}{2}-i\nu_\phi}$ at late times. 

Let us check that this is indeed the case.
Note that the connected contribution by its own contains two sets of poles associated with the first series of operators because it has both poles originating from the gamma functions in $A(\nu)$, as well as poles of $f_\sigma$. We thus expect that there is a cancellation between the poles coming from $A(\nu)$ and those of $\rho^{\text {free}}$. This condition is indeed satisfied as it turns out to be identical to the pole cancellation condition checked in \cite{DiPietro:2021sjt}. Indeed, the bubble function is the same and it has poles for the values of $\nu$ that we are interested in. Consequently the form of the propagator of $\sigma$ is not important for this check.\footnote{In fact in \cite{DiPietro:2021sjt} we had to refer to a large-$N$ theory in order to motivate the pole cancellation since one had to make sure that cross-channel diagrams do not influence the argument. Here we can ignore the connected cross channel  contributions due to large-$N$ counting. Note also that some of the poles that the bubble function has for generic $d$ are absent in even integer $d$. This, however, does not change the conclusion. }

\begin{figure}
\includegraphics[width=7cm]{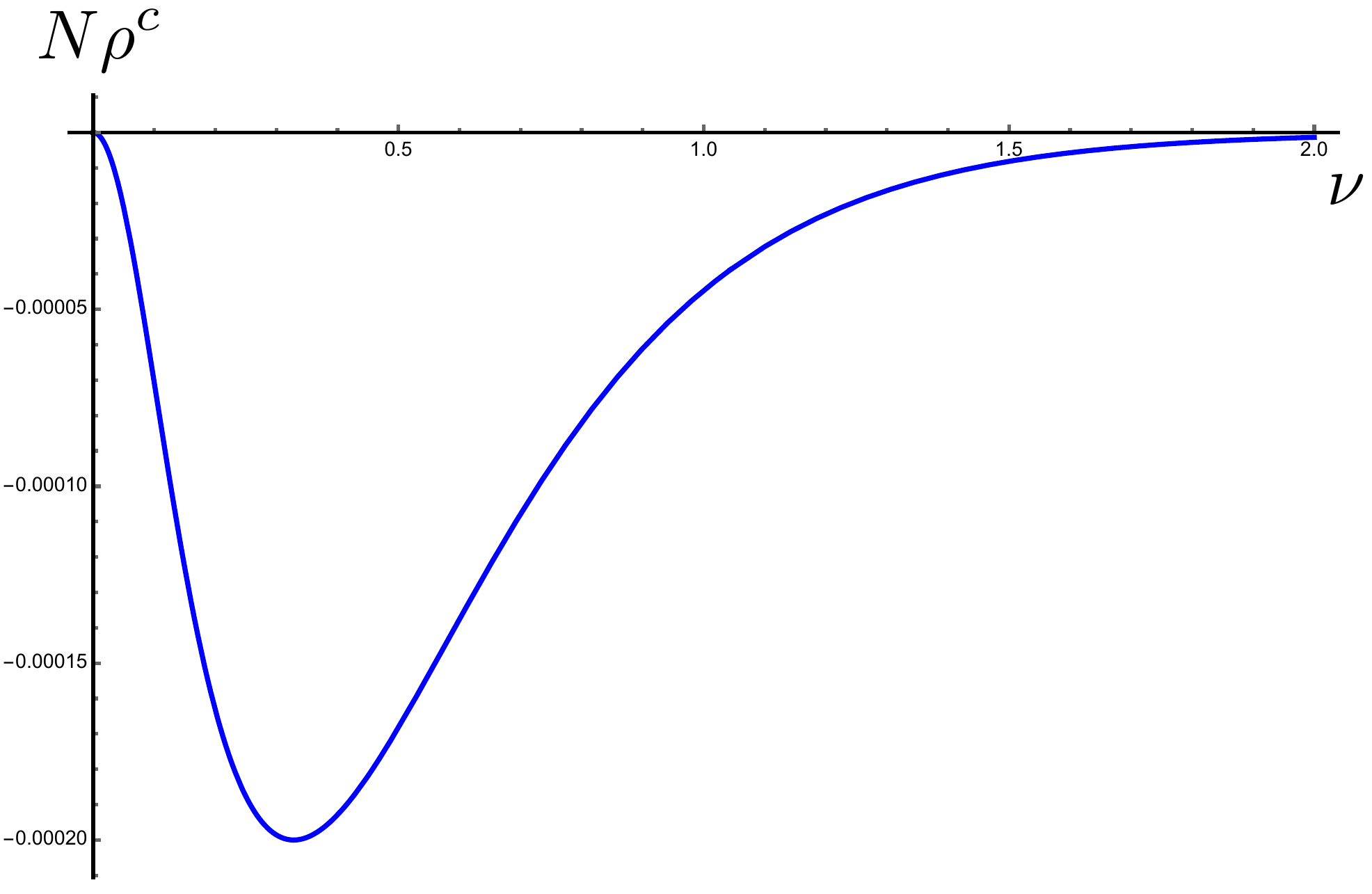}
\hspace{0.5cm}
\includegraphics[width=7cm]{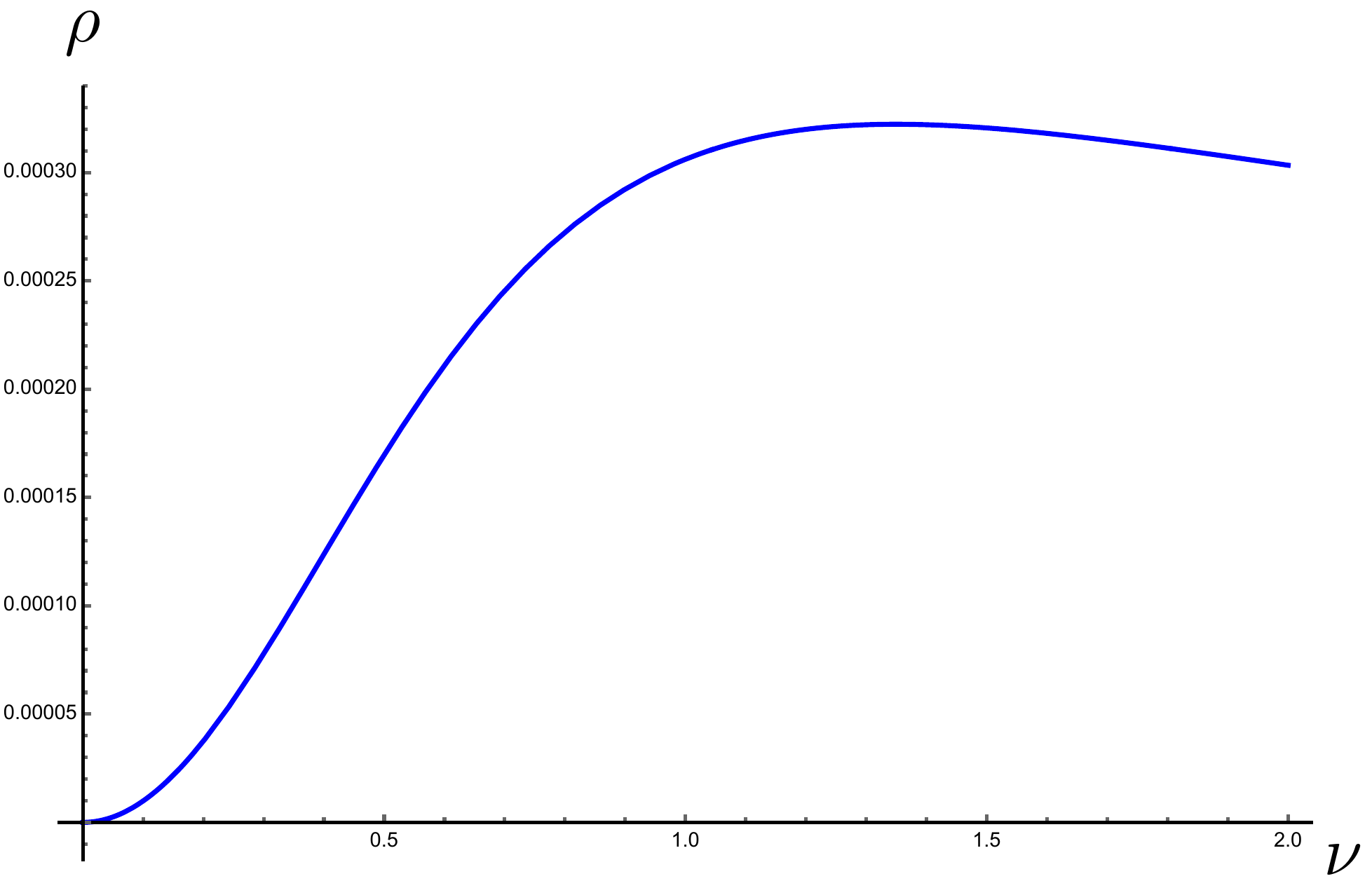}
\centering
\caption{Left: connected spectral density $N \rho^c(\nu)$ for $\lambda=20$, $\nu_\phi=-i 0.6$. Right: total spectral density $\rho(\nu)$ for the same $\lambda$ and $\nu_\phi$.} \label{fig:rho}
\end{figure} 

Let us now switch to unitarity. We restrict our attention to light external particles meaning imaginary $\nu_\phi$.\footnote{In case of heavy external particles, unitarity is more subtle since one needs to consider complex conjugate external states, as well as include contributions proportional to the delta-functions \cite{Matteo,Hogervorst:2021uvp}.} As proven in \cite{DiPietro:2021sjt,Hogervorst:2021uvp} unitarity demands that $\rho(\nu)$ is non-negative for real $\nu$, as well as that it has positive residues in case there are poles corresponding to light intermediate states, $0<i\nu<\frac{d}{2}$. We are going to check that it is indeed the case restricting to $d=2$ where the analytic expression for the bubble simplifies and we also do not need to worry about UV divergences. Even in this case we do not have a simple analytic argument that would show positivity. For example, the connected part on its own is not positive definite for all choices of the parameters, see figure~\ref{fig:rho} for an example of a negative case. Since there is no parametric separation between it and the disconnected piece, it is not obvious from the expressions \eqref{f4pt} and \eqref{rhofree} that the total density is positive. In the case of parameters as in figure~\ref{fig:rho}, the free contribution is just about two times larger than the absolute value of the connected one. This is what makes the check conceptually more interesting than in the narrow resonance case considered in \cite{DiPietro:2021sjt}, where the potentially dangerous interacting part was not suppressed only in the vicinity of the resonance. We checked that the spectral density is indeed positive by numerically scanning over  $\lambda$, $\nu_\phi$ and $\nu$. 

\section{Conclusions}\label{sec:conclusion}
In this paper, we analyzed the large $N$ $O(N)$ vector model on a rigid dS background. The large $N$ solvability of the model enabled detailed analysis of the phase structure and the late-time four-point functions. The two main lessons of our analysis are 1.~strong infrared divergences in the light scalar fields prohibit the spontaneous breaking of the continuous symmetry in this model, much like what happens in flat two-dimensional space.~2.~the late-time four-point functions of vector fields $\hat{\phi}^{i}$ exhibit meromorphicity and positivity in the spectral parameter space, the properties argued for from basic assumptions in \cite{Hogervorst:2021uvp,DiPietro:2021sjt}. In particular, positivity is restored only after combining the disconnected and the connected contributions properly, thereby providing a highly nontrivial check of unitarity.

One obvious future direction is to generalize the analysis of this paper to a broader class of solvable large $N$ theories, such as the Gross-Neveu model and Chern-Simons matter theories, and verify meromorphicity and positivity of the late-time four-point functions in the spectral parameter space.

We showed that the results from the large $N$ analysis agree with the results from the stochastic approach in their overlapping regime of validity. To make connections between the two approaches more concrete, it would be interesting to study the probability distribution of scalar fields --- the basic element in the stochastic approach --- using the large $N$ techniques. One possible approach is to study the sphere partition function with sources turned on, which will contain as much information as the probability distribution, using the large $N$ methods.

In this paper, we mostly focused on the study of dS$_3$. To study the $O(N)$ model in higher-dimensional de Sitter, one has to regularize and renormalize the ultraviolet (UV) divergence of the coupling $\lambda$. It is generally an interesting problem to study analyticity and unitarity in the presence of UV divergences. See \cite{Chowdhury:2023arc} for recent discussions on the regularization preserving the de-Sitter isometry, which is potentially useful for this purpose.

Another important direction is to find an example of spontaneous internal symmetry breaking in dS, or prove rigorously that it is never possible. A slightly related question is to understand the dynamics of compact scalars in de Sitter at the non-perturbative level, which potentially have implications on the physics of axion, see \cite{Chakraborty:2023eoq} for the analysis in perturbation theory. It is also interesting to consider higher form symmetries \cite{Gaiotto:2014kfa}, as well as confinement/deconfinement phase transitions in dS. We should note that in practical sense spontaneous symmetry breaking can occur, if the scale of symmetry breaking $f_a$ is much higher than the Hubble scale. In this case, for times $t\ll f_a^2/H^3$ and in a given Hubble volume the theory finds itself in an effective vacuum with a given value of the symmetry breaking order parameter. This also assures a smooth limit to the flat space symmetry breaking phenomena when $H\to0$. We leave a detailed investigation of this regime to future work.

An exciting future direction is to derive constraints on cosmological correlators from crossing and unitarity using the bootstrap ideology. Such procedure was already initiated in \cite{Hogervorst:2021uvp}. A potentially useful way to organize the constraints is to use the sum rules of ref. \cite{Bonifacio:2023ban} derived to bound the spectrum of the laplacian on hyperbolic manifolds \cite{Bonifacio:2021msa,Kravchuk:2021akc}. The unitarity condition in this case is the same as in dS -- positivity of the spectral density of unitary representations of the Euclidean conformal group. While this bootstrap approach can be used in principle to put constraints on genuinely strongly coupled theories in dS, it is harder to apply them to concrete perturbative setups. This is because the sum rules presented in \cite{Bonifacio:2023ban} all have a contribution of the identity, which will be dominant compared to the connected contributions, either due to a small coupling or due to large $N$, making the constraints trivial. An interesting setup in which the connected contribution is not suppressed, and these sum rules can be applied fruitfully, is that of theories in dS which produce strongly non-Gaussian correlators. One example is a theory of a single self-interacting scalar which has a small coupling constant, but at the same time light enough so that the perturbation theory leads to IR divergences. After the divergences are handled, the correlation functions are those of a Euclidean CFT with a large number of light operators, whose OPE coefficients are order one and are related to the overlaps of Eigenfunctions of a certain auxiliary differential operator \cite{Starobinsky:1986fx, Starobinsky:1994bd, Gorbenko:2019rza}. Operator dimensions are in turn related to the Eigenvalues. Such CFT data obviously fails to satisfy usual (Lorentzian) CFT unitarity due to operators of dimension below the unitarity bound, and at the same time should satisfy in a non-trivial way the sum rules.  

It would also be interesting to explore the whether non-perturbative techniques available in the $O(N)$ model are useful for studying other strongly coupled problems relevant for cosmology, for example the one studied in \cite{Celoria:2021vjw}.

\subsection*{Acknowldgement}
We thank Andreas Karch, Hayden Lee, Juan Maldacena, Bendeguz Offertaler, Guilherme Pimentel, Ivo Sachs, Pierrre Van Hove, Sasha Zhiboedov for discussions. The results of this paper were presented in the workshops, {\it Cosmology, Quantum Gravity, and Holography: the Interplay of Fundamental Concepts} in CERN, {\it Correlators in Cortona} in Palazzone, {\it DESY Theory workshop 2023} in DESY, and we thank organizers and participants for helpful discussions and feedback. LD acknowledges support from the INFN Iniziativa Specifica ST\&FI. VG acknowledges support form the SNF starting grant ``The Fundamental Description of the Expanding Universe''.

\appendix
\section{Bubble from sphere}\label{ap:bubble}
In this appendix, we explain how to compute the bubble diagram $\hat{B}^{S}(l)$ on a sphere. The same computation has been discussed in Appendix B of \cite{Marolf:2010zp} based on a slightly different approach. Here we focus on dS$_3$ ($S^3$) and perform the computation in a more straightforward way, spelling out details of the computation. For simplicity of notations, in what follows we set the radius of the sphere $L$ to be $1$.

A scalar propagator on $S^3$ can be expanded in terms of the harmonic function as
\ba
G(x,y)\langle \phi(x)\phi(y)\rangle=\sum_{l=0}^{\infty}\frac{\Omega^{S^{3}}(l,x,y)}{l(l+2)-\sigma (\sigma+2)}\comma
\ea
where $x$ and $y$ are points on $S^{3}$ and we are following the notation of \cite{Marolf:2010zp} to express the mass as $m^2=\nu^2+1=-\sigma (\sigma+2)$. Using the orthogonality of the harmonic functions, we can decompose the square of the propagator, which leads to a bubble diagram, in the following way:
\ba
\begin{aligned}
G(x,y)^2=&\sum_{n=0}^{\infty}\sum_{m=0}^{\infty}\frac{\Omega^{S^{3}}(n,x,y)\Omega^{S^3}(m,x,y)}{\left[n(n+2)-\sigma (\sigma+2)\right]\left[m(m+2)-\sigma (\sigma+2)\right]}\\
=&\sum_{n=0}^{\infty}\sum_{m=0}^{\infty}\frac{\Omega^{S^3}(l,x,y)}{\left[n(n+2)-\sigma (\sigma+2)\right]\left[m(m+2)-\sigma (\sigma+2)\right]}\\
&\times\frac{2\pi^2}{(l+1)^2}\int_{S^3}\sqrt{g(x)}\Omega^{S^3}(l,x,y)\Omega^{S^3}(m,x,y)\Omega^{S^3}(n,x,y)\period 
\end{aligned}
\ea
Performing the triple integral explicitly, we obtain
\ba
\left(G_{\nu}(x,y)\right)^2&= \sum_{l=0}^{\infty}\hat{B}^{S}(l)\Omega^{S^{3}}(l,x,y)\comma
\ea
with
\ba\label{eq:BSap}
\hat{B}^{S}(l)=\frac{2\pi^2}{l+1}\sum_{\{m,n\}}\frac{(1+m)(1+n)}{4\pi^4(m-\sigma)(2+m+\sigma)(n-\sigma)(2+n+\sigma)}\period
\ea
Here the sum $\sum_{\{m,n\}}$ is over the range,
\ba
|n-m|\leq l\leq m+n\comma\qquad  n\geq 0\comma\qquad m\geq 0\period
\ea
To perform this double sum explicitly, it is useful to change the summation variables as was done in \cite{Marolf:2010zp}:
\ba
m=G+K\comma\qquad n=G-K+l\period
\ea
In terms of the new variables, \eqref{eq:BSap} reads
\ba
\hat{B}^{S}(l)=\frac{2\pi^2}{l+1}\sum_{G=0}^{\infty}\sum_{K=0}^{l}\frac{(1+G+K)(1+G-K+l)}{4\pi^4(G+K-\sigma)(G-K+l-\sigma)(2+G+K+\sigma)(2+G-K+l+\sigma)}\period\nonumber
\ea

To further simplify the analysis, we extend the range of the summation for $K$ from $[0,l]$ to $(-\infty,\infty)$ by adding the summation ranges $[l+1,\infty)$ and $(-\infty,-1]$. Of course, one should worry if the final result remains the same after this extension. To see this, let us consider the sum for $K\in [l+1,\infty)$ by performing a redefinition $K\to K+1+l$ and expressing the sum as
\ba
\sum_{G=0}^{\infty}\sum_{K=0}^{\infty}\frac{(2+G+K+l)(K-G)}{4\pi^4(1+G+K+l-\sigma)(1+G-K+\sigma)(1-G+K+\sigma)(3+G+K+l+\sigma)}\period\nonumber
\ea
Since the summand is anti-symmetric with respect to the exchange of $G$ and $K$, one might conclude that this sum from $K\in [l+1,\infty)$ vanishes. A similar argument leads to a conclusion that the sum from $K\in (-\infty,-1]$ also vanishes. However these conclusions are incorrect: the double sum above is not absolutely convergent and the results computed in different orders of summations give different results. Fortunately, in the case at hand, there is a way to fix this problem. If we differentiate the sum by $\sigma$ or $l$, the sum becomes absolutely convergent. This implies that the ambiguity in the original sum which depends on the order of summations is simply a constant that does not depend on $\sigma$ or $l$. Therefore, in what follows, we first  perform the computation by extending the sum to $K\in (-\infty,\infty)$ and later fix the constant by computing the double sum for $l=0$ (for which the sum reduces to a single sum) more directly.

Having made these remarks, let us now perform the summation. A direct computation (by Mathematica) leads to
\begin{align}
&\sum_{G=0}^{\infty}\sum_{K=-\infty}^{\infty}\frac{(2+G+K+l)(K-G)}{4\pi^4(1+G+K+l-\sigma)(1+G-K+\sigma)(1-G+K+\sigma)(3+G+K+l+\sigma)}\nonumber\\
&=-\sum_{G=0}^{\infty}\frac{(1+\sigma)\cot (\pi \sigma)}{2\pi^3 (2G+l-2\sigma)(4+2G+l+2\sigma)}\label{eq:naivesumap}\\
&=\frac{\cot (\pi \sigma) \left(\psi(\frac{l}{2}-\sigma)-\psi(2+\frac{l}{2}+\sigma)\right)}{16\pi^3}\period\nonumber
\end{align}
Now to fix the constant, we set $l=0$ in the orignal sum \eqref{eq:BSap}. This leads to
\ba
\hat{B}^{S}(0)=\frac{1}{8 \sin (\pi \sigma)^2}-\frac{\cot (\pi \sigma)}{8\pi (1+\sigma)}\period
\ea
Comparing this with \eqref{eq:naivesumap}, we find that \eqref{eq:naivesumap} misses the constant $1/16\pi^2$. Adding it by hand, one obtains
\ba
\hat{B}^{S}(l)=\frac{1}{l+1}\frac{\pi+\cot (\pi \sigma)\left(\psi(\frac{l}{2}-\sigma)-\psi(2+\frac{l}{2}+\sigma)\right)}{8\pi}\period
\ea
After the analytic continuation, this gives $\hat{B}(\nu)$ used in the main text \eqref{eq:bubbled2}. This expression also agrees with the direct Lorentzian computation.

Let us also make a brief comment on the recent result in \cite{Polyakov:2022dpa}, which presented a different analytic expression for the bubble diagram in dS$_3$. In our notations and conventions, their result reads
\ba\label{eq:kronecker}
\hat{B}^{S}(l)=\frac{\delta_{l,{\rm even}}}{8}+\frac{\cot(\pi \sigma)\left[\psi (\frac{l}{2}-\sigma)+\psi (1-\frac{l}{2}+\sigma)-\psi(2+\frac{l}{2}+\sigma)-\psi(-1-\frac{l}{2}-\sigma)\right]}{16\pi}\comma
\ea
where $\delta_{l,{\rm even}}$ gives $1$ only for even integers and they called it a ``Kronecker anomaly'': a non-analytic dependence on the angular momentum. However this apparent non-analytic dependence is simply an artifact of the representation they chose, and for integer $l$, it coincides with our expression, which is fully analytic. The equivalence can be readily shown using the following functional identity of the digamma function:
\ba
\psi (1-x)-\psi(x)=\pi\cot (\pi x)\period
\ea
In addition, their representation \eqref{eq:kronecker} is not suited for the computation in de Sitter since it has infinitely many poles on the real non-integer values of $l$ and therefore makes it difficult to perform the Sommerfeld-Watson transformation needed for the analytic continuation from sphere to dS \cite{DiPietro:2021sjt,Marolf:2010zp}. In contrast, our representation is free of poles on the real axis of $l$ and makes it straightforward to perform the Sommerfeld-Watson transform.
\bibliographystyle{JHEP}
\bibliography{references}

\end{document}